\documentclass[aps,prb,reprint]{revtex4-1}

\usepackage[utf8x]{inputenc}
\usepackage[english]{babel}
\usepackage{amsmath}
\usepackage{amsfonts}
\usepackage{verbatim}
\usepackage[dvips]{graphicx}
\usepackage{color}
\usepackage{hyperref}
\usepackage{xspace}
\usepackage{grffile}
\usepackage{todonotes}
\usepackage{multirow}

\begin{document}

\title{Comprehensive \textit{Ab Initio} Study of Doping in Bulk ZnO with Group-V Elements}

\author{Guido Petretto}
\email{guido.petretto@cea.fr}
\affiliation{CEA, DEN, Service de Recherches de M\'etallurgie Physique, F-91191 Gif-sur-Yvette, France}

\author{Fabien Bruneval}
\email{fabien.bruneval@cea.fr}
\affiliation{CEA, DEN, Service de Recherches de M\'etallurgie Physique, F-91191 Gif-sur-Yvette, France}

\begin{abstract}
Despite the lack of reproducible experimental confirmation, group-V elements have been considered as possible sources of \textit{p}-type doping in ZnO in the form of simple and complex defects. Using \textit{ab initio} calculations, based on state-of-the-art hybrid exchange-correlation functional, we study a wide range of defects and defect complexes related with N, P, As, and Sb impurities. We show that none of the candidates for \textit{p}-type doping can be considered a good source of holes in the valence band due to deep acceptor levels and low formation energies of compensating donor defects. In addition, we discuss the stability of complexes in different regimes.
\end{abstract}

\maketitle

\section{Introduction}

Zinc oxide has attracted considerable interest as a promising material for optoelectronics applications, such as light-emitting diodes and solar cells, due to its large band gap and large exciton binding energy. The actual possibility of successfully employing ZnO in these kinds of devices relies on the ability of preparing a stable \textit{p}-type as well as an \textit{n}-type ZnO. Unfortunately, like other wide band gap semiconductors, ZnO suffers from a doping asymmetry problem and, while \textit{n}-type ZnO has been easily obtained, reliable \textit{p}-type ZnO has not been produced yet\cite{Avrutin2010}. 

One of the most common ways to overcome this difficulty has been to incorporate group-V elements, in the hope that they could substitute oxygen in ZnO crystal and should in principle provide shallow enough acceptor levels. In particular, nitrogen has played the role of the favorite dopant, owing to its atomic size close to that of oxygen. Experimentally, despite the great deal of effort devoted to the topic, only few successes in achieving \textit{p}-type ZnO have been reported with this kind of doping \cite{Tsukazaki_2004,Aoki_Hatanaka_2000,Ryu_Zhu_2000,Xiu_Yang_2005} and up to now none of them has eventually led to further development. 

From a theoretical point of view, the understanding of these results and the proposition of other ways of obtaining \textit{p}-type doping by means of density functional theory (DFT) calculations is a challenging task as well. In fact, the problem of the band gap underestimation related to the local or semilocal approximations of the exchange correlation functional is particularly relevant in the case of ZnO, with a calculated gap of only $\sim$0.8 eV. Calculations performed within these approximations have shown, for example, a transition level of 0.4 eV for N$_\textrm{O}$\cite{Park_Zhang_2002} and shallow levels for complexes of the form $X_\textrm{Zn}$-$2V_\textrm{Zn}$\cite{Limpijumnong2004}, but, in light of the severe gap underestimation, these results should be considered with care. Recently, some of these calculations have been updated with the introduction of hybrid functionals, which use an admixture of exact and local exchange, giving larger values of the band gap and providing more reliable results. According to the latest evaluation with hybrid functionals for some group-V related defects\cite{Oba2008,Lyons2009,Lany_Zunger_2010,Gallino2010,Puchala2012_2}, the results indicate that the acceptor levels are much deeper than expected from previous calculations, suggesting that these are not good candidates for achieving \textit{p}-type doping. Anyway, these calculations have been carried out with different approximations and several of the possible defects have not been considered, making it difficult to draw a uniform conclusion about the doping with group-V elements. To fill this gap, in this paper we consider the problem of ZnO doping with N, P, As, and Sb in the most relevant defect configurations by means of DFT calculations based on hybrid functionals. We show that it is unlikely to obtain good acceptor states from these elements and finally, we discuss the possibility of actually observing complex defects in doped samples.

The remainder of this article is organized as follows: in Sec. \ref{sec:comp_det} we introduce the methods used in our calculations and we discuss our approximations. In Sec. \ref{sec:simple} we begin our analysis of defects identified as promising for p-type doping in the literature, starting from simple substitutional defects X$_\textrm{O}$ and $X_\textrm{Zn}$ (X=N, P, As, Sb). In Sec. \ref{sec:complexes} we study complex defects. First, given the stability of the N$_\textrm{2}$ molecule and their importance for \textit{p}-type doping\cite{Cui_Bruneval_2010,Lambrecht_Boonchun_2013}, we analyze the (N$_\textrm{2}$)$_\textrm{O}$ and (N$_\textrm{2}$)$_\textrm{Zn}$ defects. Then, we consider complexes of the form  $X_\textrm{Zn}$-$2V_\textrm{Zn}$ and $X_\textrm{Zn}$-$V_\textrm{Zn}$, which are proved to be binding in ZnO \cite{Limpijumnong2004,Puchala2012}. In this section, to evaluate the binding energies of the complexes, the isolated Zn vacancy $V_\textrm{Zn}$ is studied as well. Finally, in Sec. \ref{sec:stability} we discuss the possibilities of actually observing these kinds of complexes, based on their binding energy.

\section{Computational details}
\label{sec:comp_det}

Our calculations are performed in the framework of plane-wave pseudopotential DFT as implemented in the Quantum ESPRESSO package\cite{Giannozzi_espresso}. Norm-conserving pseudopotentials are employed for all the atoms, with the semicore Zn 3\textit{d} states included in the valence electrons, and the energy cutoff set to 80~Ry. The exchange correlation potential is treated with the hybrid functional of Heyd, Scuseria, and Ernzerhof (HSE)\cite{Heyd_Scuseria_2003,*Heyd_Scuseria_2006}, based on the Perdew, Burke, and Ernzerhof (PBE) function where a fraction $\alpha$ of the exchange is replaced by Hartree-Fock (HF) exchange. However, the problem of choosing the $\alpha$ parameter to get accurate defect levels is still under debate and its value has been set based on several schemes\cite{Oba2008,Lany_Zunger_2010,Ramprasad2012}. Because we are dealing with transition energies that span all the band gap, we set $\alpha=0.45$. This allows us to correctly reproduce the experimental value of the band gap of 3.44 eV and gives lattice parameters $a=3.23$ {\AA}, $c=5.19$ {\AA} and $u=0.38$, in good agreement with the experiments. Such a large value of $\alpha$ may affect the transition levels due to the downshift of the valence band maximum (VBM) when the $\alpha$ parameter is increased, but this fact should not alter the general conclusions drawn in the following. The calculations are carried out in a 72 atom supercell with a $2\times2\times2$ Monkhorst-Pack grid \cite{Monkhorst_Pack_1976} to sample the Brillouin zone, while the Fock exchange potential is calculated using a coarser $1\times1\times1$ grid of $Q$-points\cite{Paier_Marsman_2006}. The formation energies of defect $D$ in charge state $q$ are calculated according to the formula \cite{Zhang_Northrup_1991}
\begin{multline}
\label{eq:form_en}
E_f(D,q)=E_{\textrm{tot}}(D,q)-E_{\textrm{tot}}(\textrm{bulk})\\
-\sum_i n_i\mu_i+q(\epsilon_v+\epsilon_F+\Delta V)+\Delta E_\textrm{el}(q),
\end{multline}  
where $E_{\textrm{tot}}(D,q)$ and $E_{\textrm{tot}}(\textrm{bulk})$ represent the total energy of the supercell with and without the defect, $\epsilon_v$ is the energy of the VBM, $\epsilon_F$ indicates the value of the Fermi level inside the band gap, and $\Delta V$ is a potential alignment term\cite{Taylor_Bruneval_2011}. The electrostatic correction $\Delta E_\textrm{el}(q)$ is calculated as the monopole Madelung term\cite{Leslie_Gillan_1985} in the generalized case when anisotropy in the screening requires that the dielectric constant $\varepsilon$ is replaced by a tensor $\bar{\varepsilon}$\cite{Rurali_Cartoixa_2009}:
\begin{multline}
\label{eq:mad_corr}
 \alpha = \sum_{ {\bf R} \neq {\bf 0}} \frac{1}{\sqrt{ \textrm{det } \bar{\varepsilon}} }
   \frac{\textrm{erfc}( \gamma \sqrt{ {\bf R} \cdot \bar{\varepsilon}^{-1} \cdot {\bf R} } )}{\sqrt{ {\bf R} \cdot \bar{\varepsilon}^{-1} \cdot {\bf R} }} + \\
   \sum_{ {\bf G} \neq {\bf 0}}
    \frac{4 \pi}{V_c} \frac{ \exp( -{\bf G} \cdot \bar{\varepsilon} \cdot {\bf G} / 4 \gamma^2 ) }{ {\bf G} \cdot \bar{\varepsilon} \cdot {\bf G} } -
   \frac{2 \gamma}{ \sqrt{ \pi \textrm{det } \bar{\varepsilon}}  } - \frac{\pi}{V_c \gamma^2} ,
\end{multline}
where the sum over ${\bf R}$ and ${\bf G}$ extends over the direct and reciprocal lattice vectors, respectively, except for zero. In the case of ZnO $\bar{\varepsilon}=\textrm{diag}(\varepsilon_\perp,\varepsilon_\perp,\varepsilon_\parallel)$ with $\varepsilon_\perp=7.77$ and $\varepsilon_\parallel=8.91$\cite{Ashkenov_Mbenkum_2003}. Finally, for each specie $i$, $n_i$ is the change in the total number of atoms to create defect $D$ and $\mu_i$ is the corresponding chemical potential, which are constrained by the growth conditions. In particular, $\mu_\textrm{Zn} = \mu_\textrm{Zn}^0+\Delta H_f$  and $\mu_\textrm{O}=\mu_{\textrm{O}_2}/2$ for an O-rich environment, while $\mu_\textrm{Zn} = \mu_\textrm{Zn}^0$ and $\mu_\textrm{O}=\mu_{\textrm{O}_2}/2+\Delta H_f$ for Zn-rich conditions. Here $\Delta H_f=-2.96$~eV is the calculated ZnO enthalpy of formation. If needed, also the chemical potentials of the group-V dopants should be constrained as well. The results shown are under the dopant rich condition, using molecular N$_2$ and P$_4$O$_{10}$, and solid As$_2$O$_3$ and Sb$_2$O$_3$ as dopant sources. Note that the different compositions of these sources, which include oxygen in different percentages, make a numerical comparison of the formation energies of the defects not straightforward.

\begin{figure}
\begin{center}
 \includegraphics[width=0.45\textwidth]{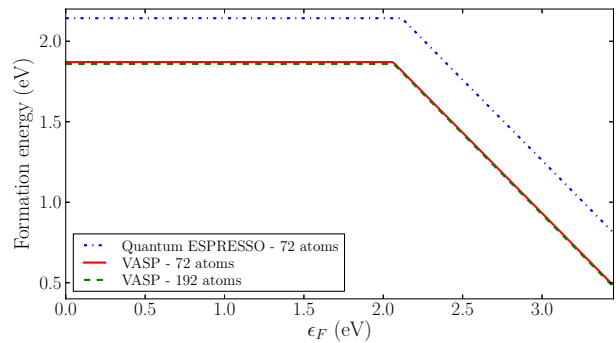}
 \caption{Formation energy $E_f$ of N$_\textrm{O}$ as a function of the Fermi level under Zn-rich conditions, as obtained from different codes and supercell sizes. \label{fig:convergence}}
\end{center}
\end{figure}

In order to assess the correctness of our approximations, we performed a test for the notable case of N$_\textrm{O}$. To this aim we used the HSE hybrid functional and projector augmented wave (PAW)\cite{Blochl_1994} potentials as implemented in the VASP code\cite{VASP_1,*VASP_2}, using a cutoff of 400 eV and setting $\alpha=0.375$ to correctly reproduce the experimental band gap\cite{Oba2008}. We considered 72 atom and 192 atom supercells, with, respectively, a $2\times2\times2$ and $\Gamma$ $k$-points sampling for the Brillouin zone. A Madelung correction calculated from Eq. \eqref{eq:mad_corr} has been applied to the charged system. As can be seen from the results in Fig. \ref{fig:convergence}, a good convergence is achieved for the neutral and charged state, confirming that a 72 atom supercell is enough to correctly describe these deep defects. In addition, this comparison assesses the reliability of the norm-conserving pseudopotentials compared to the more accurate PAW scheme. Note the impressive agreement for the transition level $\epsilon (0/-)$ and the fair agreement for the formation energy $E_f$.

Since the N$_\textrm{O}$ defect has been already widely studied, even with hybrid functionals, it is worth comparing our results with those present in the literature in the case of $\alpha$ being tuned to match the experimental band gap.The data are summarized in Table \ref{tab:N_O_data}. Our value for the $\epsilon (0/-)$ transition energy of 2.1 eV is the same as obtained by Lany and Zunger\cite{Lany_Zunger_2010} for $\alpha=0.38$. The value of 1.8 eV for $\alpha=0.375$ calculated by Boonchun and Lambrecht\cite{Boonchun_Lambrecht_2013} is very close to our value without the Madelung correction. This is also in agreement with our results obtained with VASP (see Fig. \ref{fig:convergence}). At variance, Lyons \textit{et al.}\cite{Lyons2009} and Sakong \textit{et al.}\cite{Sakong_Gutjahr_2013} find a transition energy of 1.3 and 1.46, respectively, setting $\alpha=0.36$. This is in agreement with the result of Gallino \textit{et al.}\cite{Gallino2010}, which, however, is obtained with the B3LYP hybrid functional\cite{B3LYP}, making the comparison less straightforward. The reason for the disagreement between the two groups of results is still unclear.

\begin{table}
\begin{ruledtabular}
\begin{tabular}{lccccccc}
 & \multicolumn{2}{c}{This work} &  \multicolumn{5}{c}{Ref.}\\
 & QE & VASP & \onlinecite{Lany_Zunger_2010} & \onlinecite{Boonchun_Lambrecht_2013} &  \onlinecite{Lyons2009} &  \onlinecite{Sakong_Gutjahr_2013} &  \onlinecite{Gallino2010} \\
\hline
$\alpha$ & 0.45 & 0.375 & 0.38 & 0.375 & 0.36 & 0.36 & 0.2 \\
$\epsilon (0/-)$ w/o FSC & 1.87 & 1.88 &  & 1.8 & 1.3 & 1.32 & \\
$\epsilon (0/-)$ w/ FSC & 2.10 & 2.06 & 2.1 &  &  & 1.46 & 1.47
\end{tabular}
\end{ruledtabular}
 \caption{Comparison of $\epsilon (0/-)$ transition energies for N$_\textrm{O}$. The calculations are performed with hybrid functionals and the percentage of exact exchange $\alpha$ is tuned to match the band gap with the experimental value. QE and VASP columns contain our results calculated from Quantum ESPRESSO and VASP. The data are shown, when available, both with and without finite size correction (FSC). All values are in eV. \label{tab:N_O_data}}
\end{table}

\section{Simple substitutional defects}
\label{sec:simple}

\begin{figure}
\begin{center}
 \includegraphics[width=0.45\textwidth]{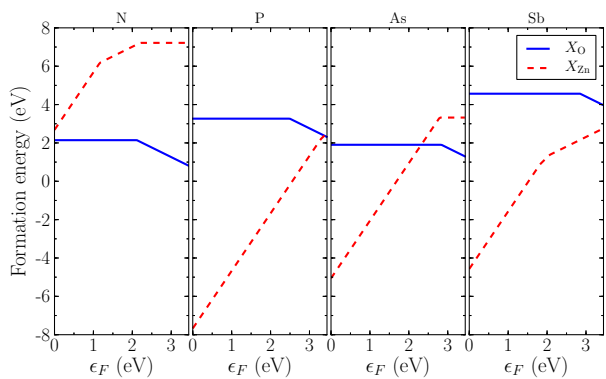}
 \caption{Formation energy $E_f$ of simple substitutional defects as a function of the Fermi level in  Zn-rich conditions. The Zn-rich conditions are chosen so to best stabilize the oxygen substitution. The zero of the Fermi level has been set to the valence band maximum. \label{fig:form_energy_Zn-rich}}
\end{center}
\end{figure}

\begin{figure}
\begin{center}
 \includegraphics[width=0.45\textwidth]{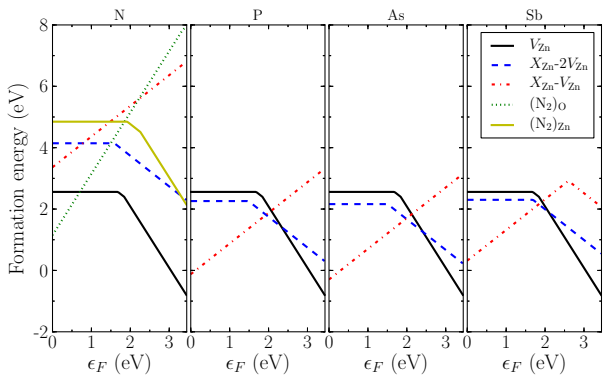}
 \caption{Formation energy $E_f$ of complex defects as a function of the Fermi level under O-rich conditions. \label{fig:form_energy_O-rich}}
\end{center}
\end{figure}

To discuss the relative stability of the different kinds of defects and the charge transition energies, we report the formation energies $E_f$ as a function of $\epsilon_F$ in Figs. \ref{fig:form_energy_Zn-rich} and \ref{fig:form_energy_O-rich}. The former figure deals with the simple substitutional defects under Zn-rich growth conditions and the latter figure focuses on complexes under O-rich conditions. As it is immediately evident, in most cases, defects of the same kind display the same qualitative behavior for all the dopants, and for P, As, and Sb a quantitative agreement can be recognized. We thus begin analyzing each specie of defect. Simple oxygen substitutional defects $X_\textrm{O}$ are single acceptors due to the missing electron in group-V elements compared to O. From a structural point of view, when using hybrid functionals the hole present in the neutral charge state is localized on one of the four bonds, producing a nonsymmetric configuration. 

It has been shown that N$_\textrm{O}$ with the hole localized along the bond parallel to the $c$ axis (N$_{\textrm{O}\parallel}$) is the most stable configuration and gives good agreement with electron paramagnetic resonance (EPR) measurements\cite{Lany_Zunger_2010,Gallino2010}. Our calculations confirm these results, with N$_{\textrm{O}\parallel}$ configuration having a 45 meV lower energy than the one with the hole localized along a bond perpendicular to the $c$ axis (N$_{\textrm{O}\perp}$), highlighting the ability of hybrid functionals to correctly describe these kinds of defects. For P$_\textrm{O}$, As$_\textrm{O}$, and Sb$_\textrm{O}$ we found the same tendency to hole localization, with the difference that $X_{\textrm{O}\parallel}$ is unstable and only configurations of the type $X_{\textrm{O}\perp}$ are observed and with a much smaller asymmetry in the bond lengths. Although defects of this kind have a stable negative charge state configuration, they fail to be good sources of holes. In fact, with the notable exception of N$_\textrm{O}$, all have quite high formation energy compared to other species, even in the Zn-rich limit, especially if the Fermi level is close to the VBM. Even if this kind of defect could be stabilized, the thermal transition energies $\epsilon (0/-)$ are 2.10, 2.58, 2.92, and 3.37 eV for N, P, As, and Sb, respectively, making them extremely deep acceptors. This should be enough to disregard the $X_\textrm{O}$ configurations as a possible source of \textit{p}-type doping.

When substituting Zn atoms, group-V elements are instead triple donors. According to our hybrid functional results, P and Sb bear positive charge state for all the values of $\epsilon_F$ making them good donors and in agreement with some experimental results \cite{Liu_Izyumskaya_2012}, while N and As has rather deep transition energies $\epsilon(0/+)=1.1$ eV and $\epsilon(0/3+)=0.73$ eV from the top of the conduction band, respectively. However, in this context it is important that, when $\epsilon_F$ is close to the valence band maximum, $X_\textrm{Zn}^{3+}$ have a very small formation energy and therefore will certainly contribute to compensate the effect of acceptor impurities that one wants to create. 

Also in this case, due to its smaller atomic radius, N$_\textrm{Zn}$ has a different structure compared to P, As, and Sb. The zinc site with four bonds is not stable and the N atom tends to shift, bonding with a lower number of oxygen atoms. We observed several inequivalent metastable configurations, whose relative stability could depend also on the charge state of the system. These configurations often include the formation of NO$_n$ ($n$=1,2,3) molecules which are isolated or loosely bound to the crystal lattice. In Fig. \ref{fig:form_energy_Zn-rich} we show only the formation energy for the most stable configurations found. For the neutral charge state N is bound to only 2 O atoms, which breaks part of the bonds with their Zn nearest neighbors. In the 1+ charge state we have a similar configuration, with one of the O breaking all the bonds, resulting in a NO$_2$ molecule bonded to the lattice through an O. Finally, for 2+ and 3+ charge states the defect results in a NO$_3$ molecule bounded to one of the neighboring Zn atoms. The final structure of the N$_\textrm{Zn}$ defect leads to strong distortions in the lattice and this will explain the much larger formation energy of N$_\textrm{Zn}$ with respect to N$_\textrm{O}$, when compared with the behavior of the other group-V elements. In fact, except for As$_\textrm{Zn}^0$, where As is strongly bound to just three O atoms, P, As, and Sb form a symmetric configuration after relaxation.

\section{Complexes}
\label{sec:complexes}

\begin{figure}
\begin{center}
 \includegraphics[width=0.45\textwidth]{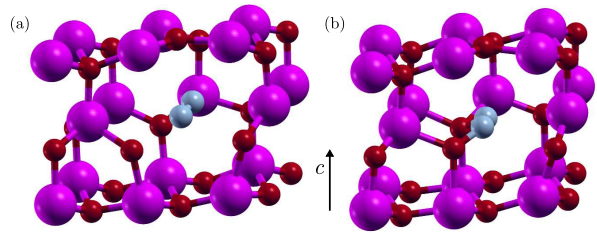}
 \caption{Possible configurations for the N$_2$ molecule on Zn site in 0 charged state. (a) The isolated configuration: the molecule is isolated from the neighboring O atoms. (b) The bridging configuration: the molecule is bonded to two O atoms. N, Zn, and O are shown as gray, violet and red spheres, respectively.\label{fig:N2_config}}
\end{center}
\end{figure}

Since we have just shown that the isolated dopant are ineffective for \textit{p}-type doping, let us now move to analyze complex defects, starting from the N$_\textrm{2}$ molecule, which can substitute an O or a Zn atom as recently identified in Ref. \onlinecite{Lambrecht_Boonchun_2013}. In the former case each N atom binds with two surrounding Zn and the molecule acts as a strong double donor. It is also quite stable, compared with N$_\textrm{O}$ and even N$_\textrm{Zn}$, and this provides a further reason for the difficulty of obtaining \textit{p}-type doping from N\cite{Cui_Bruneval_2010}. On the other hand, (N$_\textrm{2}$)$_\textrm{Zn}$ could be a double acceptor. Recently, Lambrecht and Boonchun\cite{Lambrecht_Boonchun_2013} have studied the (N$_\textrm{2}$)$_\textrm{Zn}$ molecule in an isolated configuration, excitingly concluding that it has a relatively shallow transition energy $\epsilon (0/-)$ for both PBE and HSE calculations. According to our results, however, the system has two competing configurations, one with the N$_\textrm{2}$ forming a bridge across two O atoms and one in the N$_\textrm{2}$ isolated molecule configuration with a $S$=1 spin configuration, as shown in Fig. \ref{fig:N2_config}. We found that the most stable configuration among the two is charge state dependent. In particular, for the neutral defect the bridging configuration is more stable by 0.25 eV. The opposite is true for the $1-$ and $2-$ charge states, where the isolated configuration is much more favorable. Anyway, in both cases the configuration with higher energy is metastable. The same trends are observed within PBE calculations, even if the formation energy differences are smaller in this case. With our calculation parameters, both the configurations in the neutral charge state have relatively low formation energies, making the transition level very deep ($\epsilon (0/-)=1.92$ eV), at variance with what has been observed in Ref. \onlinecite{Lambrecht_Boonchun_2013}. In addition, (N$_\textrm{2}$)$_\textrm{Zn}$ is also less stable than (N$_\textrm{2}$)$_\textrm{O}$, even in O-rich conditions, and thus unlikely to be a good acceptor.

It should be noted that the neutral isolated configuration is in a spin polarized state with spin $S$=1, as it happens for the V$_\textrm{Zn}$ defect. This, along with the similar behavior for the transition energies of (N$_\textrm{2}$)$_\textrm{Zn}$ and V$_\textrm{Zn}$, could suggest that the N$_\textrm{2}$ molecule remains quite inert with respect to the environment and that the (N$_\textrm{2}$)$_\textrm{Zn}$ defect acts basically as a Zn vacancy.

Given the difficulties of finding a shallow acceptor among simple donors configurations, complexes of the form $X_\textrm{Zn}$-$2V_\textrm{Zn}$ are studied, since they are expected to be stable acceptors due to reaction $X_\textrm{Zn}^{3+}+2V_\textrm{Zn}^{2-} \rightarrow  (X_\textrm{Zn}-2V_\textrm{Zn})^{-}$. Since the first time they were proposed \cite{Limpijumnong2004}, due to the results of semilocal functional calculations \cite{Li2011} these complexes are believed to be shallow acceptors and have often been used to justify the observation of \textit{p}-type doping in experiments. Only recently Puchala and Morgan \cite{Puchala2012_2} have shown that As$_\textrm{Zn}$-$2V_\textrm{Zn}$ is instead a deep acceptor according to hybrid DFT calculations. 

\begin{figure}
\begin{center}
 \includegraphics[width=0.45\textwidth]{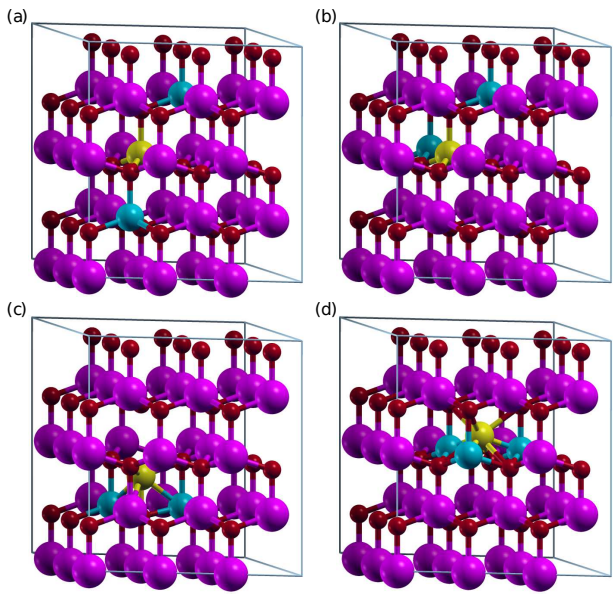}
 \caption{The four different starting configurations considered for the $X_\textrm{Zn}$-$2V_\textrm{Zn}$ defects (see text). $X$, Zn, and O are shown as yellow, violet, and red spheres, respectively, while the blue spheres represent the removed Zn atoms.\label{fig:X-2V_config}}
\end{center}
\end{figure}

Several inequivalent configurations can be found removing three Zn atoms and adding one impurity. We thus consider four of them, shown in Fig. \ref{fig:X-2V_config}, with the following different qualitative peculiarities: (1) the two vacancies placed at opposite sites with respect to the dopant\cite{Limpijumnong2004} [Fig. \ref{fig:X-2V_config}(a)], (2) two neighboring vacancies next to the dopant\cite{Li2011} [Fig. \ref{fig:X-2V_config}(b)], and two neighboring vacancies with the dopant shifted in (3) tetrahedral [Fig. \ref{fig:X-2V_config}(c)] and (4) octahedral [Fig. \ref{fig:X-2V_config}(d)] interstitial positions\cite{Puchala2012_2}. For N the most stable structure is obtained starting from configuration (3), but it ends up in a distorted configuration, with N bound to only three O atoms. When the dopant is P or As configurations (2) and (3) are the most favorable and almost equivalent in energy, with (2) being slightly favored. Sb, having a larger size, prefers configuration (4). 

Here only the values of these most stable configurations are reported. It can be argued that the small sizes of the supercell could play a role in determining the most favorable configuration, since the complex is quite extended and this could actually be the case, as demonstrated in some cases \cite{Puchala2012}. However, the differences in the formation energies between the various configurations are usually below few hundreds of meV, a quantity that will only marginally affect the values of formation and binding energies and will not alter the relevant conclusions of our work, like the transition energies being too deep and the formation energies of $X_\textrm{Zn}$-$2V_\textrm{Zn}$ defects being much larger than those of $X_\textrm{Zn}$-$V_\textrm{Zn}$ close to the VBM. 

These defects require several missing Zn, therefore the O-rich conditions should make them more stable. It can be seen in Fig. \ref{fig:form_energy_O-rich} that, in this limit, $X_\textrm{Zn}$-$2V_\textrm{Zn}$ is the acceptor with the lowest formation energy, even if still larger than that of $X_\textrm{Zn}$ for a wide range inside the band gap. Despite their relatively low formation energies, we are faced again with the problem of deep values of the ionization energies $\epsilon(0/-)$ for all the group-V elements: 1.53, 1.48, 1.50, and 1.69 eV for N, P, As, and Sb, respectively. Unlike for $X_\textrm{O}$ defects, the transition energies are approximately the same for all $X$ and much smaller, but still too deep to provide good acceptor levels.

Because $X_\textrm{Zn}$-$2V_\textrm{Zn}$ are complex defects that involve three atoms, it is necessary to study the properties of its simpler components, namely $X_\textrm{Zn}$-$V_\textrm{Zn}$ and $V_\textrm{Zn}$. Complexes with only one vacancy X$_\textrm{Zn}$-$V_\textrm{Zn}$ are quite stable as well, but they do not undergo strong deformation. They are just hold together by the attractive Coulomb interaction between $X_\textrm{Zn}^{3+}$ and $V_\textrm{Zn}^{2-}$ and, as expected, they are donors. 

Once again the presence of the N$_\textrm{Zn}$ leads to quite large formation energies for N$_\textrm{Zn}$-$2V_\textrm{Zn}$ and N$_\textrm{Zn}$-$V_\textrm{Zn}$. This should be considered as a direct consequence of the large formation energy of N$_\textrm{Zn}$ and of the lattice distortions that come with it. 

Finally, it can be seen that zinc vacancy is a double acceptor, but with a deep transition level $\epsilon (0/-)=1.68$ eV and a high formation energy close to the VBM. As mentioned before, the neutral charge state is more stable in a spin polarized configuration, as in the case of the isolated molecule configuration of (N$_\textrm{2}$)$_\textrm{Zn}$.

\section{Stability of  vacancy related complexes}
\label{sec:stability}

\begin{figure}
\begin{center}
 \includegraphics[width=0.45\textwidth]{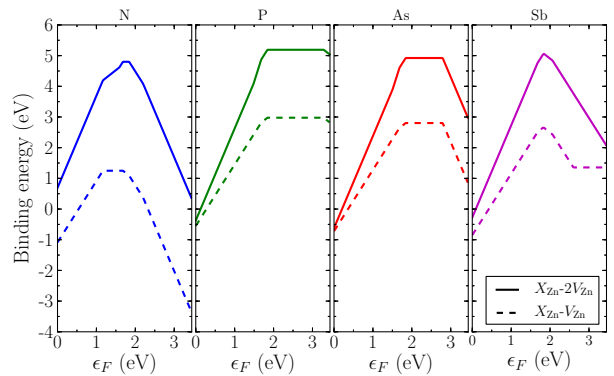}
 \caption{Binding energy $E_b$ of complex defects as a function of the Fermi level. \label{fig:bind_energy}}
\end{center}
\end{figure}

Up to now, we have shown that these complexes should not be suitable to create \textit{p}-type ZnO. Nonetheless, there are some speculations about their possible presence in experiments\cite{Azarov_Knutsen_2013,SenthilKumar2013} and so it is interesting to check whether these kinds of defects are likely to be found or not. For the cluster to form, it is essential that their binding energy $E_b=E_f(X)+nE_f(V_\textrm{Zn})-E_f(X\textrm{-}nV_\textrm{Zn})$ is positive and large enough to favor the complexes over the single components. The values of $E_b$ for all the kinds of dopants are shown in Fig. \ref{fig:bind_energy} as a function of the Fermi level. The complexes including N are much less binding than those for the other group-V elements, and, in particular, N$_\textrm{Zn}$-V$_\textrm{Zn}$ is almost never binding. As a further confirmation of the scarce utility of the complexes we can see that they are just loosely binding in the \textit{p}-type regime, while the binding energy grows rapidly with $\epsilon_F$ and could remain as large as 5 eV up to the conduction band minimum for P$_\textrm{Zn}$-2V$_\textrm{Zn}$. This suggests that it should be possible to observe them for large enough values of $\epsilon_F$.

\begin{figure}
\begin{center}
 \includegraphics[width=0.45\textwidth]{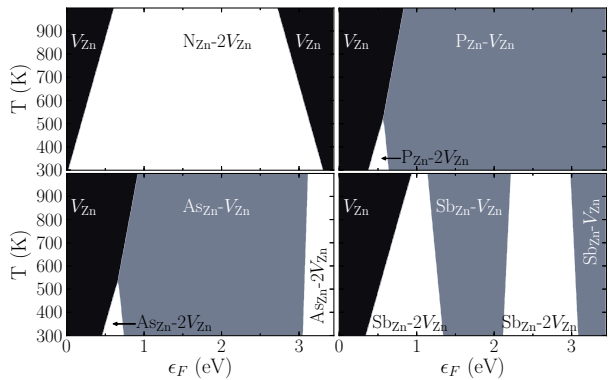}
 \caption{Phase diagram showing the prevalent configuration of the zinc vacancy configurations as a function of temperature and Fermi level. Black $V_\textrm{Zn}$, gray $X_\textrm{Zn}$-$V_\textrm{Zn}$ and white $X_\textrm{Zn}$-$2V_\textrm{Zn}$.\label{fig:defect_phase_diag}}
\end{center}
\end{figure}

To further investigate this possibility, we consider the case when the defects are formed out of equilibrium in the crystal during the growth and the total amounts of dopants and vacancies remain fixed during the cooling down of the sample. In this process the defects could reorganize in complexes and the evolution of the concentrations as function of the temperature $T$ is governed by the mass-action law
\begin{equation}
\label{eq:mass_action}
[X\textrm{-}nV_\textrm{Zn}]=[X][V_\textrm{Zn}]^n\exp(E_b/k_BT), \qquad n=1,2
\end{equation}
where [$X$-$n$V$_\textrm{Zn}$], [$X$], and [$V_\textrm{Zn}$] are the defect concentrations and $k_B$ is the Boltzmann constant. The experimentally reported densities of group-V elements are quite variable in the range $10^{15}$--$10^{20}$ cm$^{-3}$, while Zn vacancies\cite{Tuomisto_Ranki_2003} could be found with a density of $10^{15}$--$10^{16}$ cm$^{-3}$. Thus we fix the total concentrations to [$X$]$_\textrm{tot}$=$1\times10^{17}$ cm$^{-3}$ and [$V_\textrm{Zn}$]$_\textrm{tot}$=$1\times10^{16}$ cm$^{-3}$, where 
\begin{equation}
\label{eq:constraints}
\begin{array}{r@{}l}
[\textrm{X}]_\textrm{tot} &{} = \sum_n[\textrm{X}-n\textrm{V}_\textrm{Zn}], \\ [\medskipamount]

[\textrm{V}]_\textrm{tot} &{} = [\textrm{V}_\textrm{Zn}] + \sum_n n[\textrm{X}-n\textrm{V}_\textrm{Zn}].
\end{array}
\end{equation}
Solving the system given by Eqs. \eqref{eq:mass_action} and \eqref{eq:constraints} for $E_b(\epsilon_F)$ and $T$ down to room temperature, we verified that the concentrations tend to saturate quite fast with one of the defect among $X_\textrm{Zn}$-$2V_\textrm{Zn}$, $X_\textrm{Zn}$-$V_\textrm{Zn}$, and $V_\textrm{Zn}$ dominating the others. Of course, in our configuration the limiting factor for complex total concentration is [$V_\textrm{Zn}$]$_\textrm{tot}$, while the group-V elements are supposed to be present in abundance. We have verified that changes of a few orders of magnitude in the values [$X$]$_\textrm{tot}$ and [$V_\textrm{Zn}$]$_\textrm{tot}$ only lead to small changes in the relative concentrations and thus do not alter qualitatively our findings.

The phase diagrams reporting the defect with the highest concentration are shown in Fig. \ref{fig:defect_phase_diag} and they confirm that in the $p$-type regime the Zn vacancy should be the dominant defect, ruling out the $X_\textrm{Zn}$-$2V_\textrm{Zn}$ as doping sources. For larger values of $\epsilon_F$, instead, it appears that both $X_\textrm{Zn}$-$2V_\textrm{Zn}$ and $X_\textrm{Zn}$-$V_\textrm{Zn}$ could be obtained, in agreement with some experimental result.  As a final remark, it should be pointed out that the dependence of the defects diffusivity on $\epsilon_F$ and $T$\cite{Janotti2007,Puchala2012} could hinder the reorganization of the defects into complexes when decreasing too quickly the temperature.

\section{Conclusion}
\label{sec:conclusion}

In conclusion, we studied the properties of ZnO doped with group-V elements as possible candidates to provide \textit{p}-type doping. We show that none of the defect configurations considered so far are likely to provide a good source of holes, due to the defects being too deep or even worse to their tendency to behave like donors. We consider the complexes $X_\textrm{Zn}$-$2V_\textrm{Zn}$ and $X_\textrm{Zn}$-$V_\textrm{Zn}$ and we provide a justification to the experimental observation of these kinds of defects.

\begin{acknowledgments}
We thank W. R. L. Lambrecht and V. Sallet for valuable discussions. This work was supported by the French ``Agence Nationale de la Recherche'' (Project No. ANR-11-NANO-013). This work was performed using HPC resources from GENCI-IDRIS and GENCI-TGCC (Grant 2013-gen6018).
\end{acknowledgments}

\end{document}